\documentclass[aps,eqsecnum,amsmath,onecolumn,groupedaddress,superscriptaddress,notitlepage,nofootinbib]{revtex4-1}

\usepackage[utf8]{inputenc}
\usepackage{indentfirst}
\usepackage{amssymb}
\usepackage{amsmath,latexsym}
\usepackage{graphicx}
\usepackage{float}
\usepackage[margin=1.0in]{geometry}
\usepackage[mathscr]{euscript}
\usepackage{color}
\usepackage{xcolor}
\usepackage{soul}
\usepackage{slashed}
\graphicspath{{images/}}
\usepackage[mathscr]{euscript}
\usepackage[force]{feynmp-auto}
\usepackage{subcaption}
\DeclareMathOperator{\Tr}{Tr}

\newcommand{\beq}{\begin{equation}}
\newcommand{\eeq}{\end{equation}}
\newcommand{\bea}{\begin{eqnarray}}
\newcommand{\eea}{\end{eqnarray}}
\newcommand{\e}{\text{e}}
\def\beqs#1\eeqs{\beq\begin{split} #1 \end{split}\eeq}

\usepackage[colorlinks=true,backref=false, linktocpage=true,
citecolor=red,urlcolor=red,linkcolor=red,pdfpagemode=UseOutlines]{hyperref}

\hypersetup{%
  bookmarksnumbered=true,
  pdftitle = {},
  pdfsubject = {},
  pdfauthor = {},
  pdfkeywords = {}
}

\usepackage{microtype}

\newcommand{\nn}{\nonumber}
\newcommand{\eq}[1]{Eq.~(\ref{#1})}
\newcommand{\fig}[1]{Fig.~\ref{#1}}
\newcommand{\tab}[1]{Table~\ref{#1}}

\def\fm {\,{\tt fm}}
\def\MeV {\,{\tt MeV}}

\begin{document}
\title{Structure Factors of The Unitary Gas Under Supernova Conditions}

\author{Andrei Alexandru}
\email{aalexan@gwu.edu }
\affiliation{Department of Physics, The George Washington University,
Washington, DC 20052}
\affiliation{Department of Physics,
University of Maryland, College Park, MD 20742}
\author{Paulo F. Bedaque}
\email{bedaque@umd.edu}
\author{Neill C. Warrington}
\email{ncwarrin@umd.edu}
\affiliation{Department of Physics,
University of Maryland, College Park, MD 20742}

\date{\today}
\begin{abstract}
We compute with lattice field theory the vector and axial static structure factors of the unitary gas for arbitrary temperature above the superfluid transition and for fugacities $0.1 < z < 1.0$.  Using the lattice formulation, we calculate beyond the validity of the virial expansion, a commonly used technique in many-body physics. We find qualitative differences in the behavior of the structure factors at high fugacity compared to the predictions of the virial expansion. Due to the large scattering length of neutrons, we expect the unitary gas structure factors to approximate the structure factors of hot neutron gases, and we therefore expect our calculations to be useful in supernova simulations, where neutron gas structure factors are needed to compute in-medium neutrino-neutron scattering rates.
\end{abstract}
\maketitle

\section{Introduction}
The unitary gas, a gas of spin $1/2$ fermion with short range interactions tuned so a two-body bound state exists at threshold, is a classic example of a conformal system. The challenge in studying it stems from the fact that it is a strongly coupled theory for which no perturbative method applies. At the same time, a number of physical systems in Nature, varying over a large range of energy scales, are very close to being a unitary gas. Besides dilute atomic fermionic gases in atomic traps, the dilute and warm neutron gas, as found in core collapse supernova, is close to being a unitary gas. Indeed, the conditions required for a neutron gas to be close to the unitary gas is that 
the typical interparticle distance $l$ should lie between the scattering length $a$ and the range of the forces $R$, while, at the same time, the thermal wavelength $\lambda_T$ should be larger than $R$, so that the details of the potential are not probed during collisions.
For a neutron gas where the scattering length is $a \simeq -23 \fm$ and the range of the nuclear forces $R$ set by the inverse pion mass , $R\approx 1/m_\pi \approx 1.3 \fm$, the neutron gas will approximate well a unitary gas over the range of densities and temperatures
$0.01 n_0 \leq n \leq 0.1 n_0$ ($n_0\approx 0.16/\fm^3$ is the  density of nuclear matter) and $ T \leq 10 \MeV$. This is similar to the range of densities and temperatures found in the neutrinosphere of core collapse supernovae~\cite{PhysRevC.96.055804,Horowitz:2005nd,HOROWITZ2006153}, the region where neutrinos decouple from  matter during the explosion, and which is essential to the modeling of the core collapse supernova process~\cite{RevModPhys.85.245,doi:10.1146/annurev.nucl.55.090704.151608}.

The interactions between the neutrinos and the neutron matter are controlled not only by the cross section of neutrinos scattering off individual neutrons, but also by many-body effects. As a particularly dramatic demonstration of this, recent 3D supernova simulations have shown
that small changes in the scattering rate, of the order that could be caused by in-medium effects~\cite{Melson_2015}, can turn a dud into an explosion.
The in-medium differential scattering rate of neutrinos due to neutral current scattering processes in the elastic (zero energy transfer) approximation is given by~\cite{HOROWITZ2006326}:
\beq
\frac{d\Gamma}{d\cos\theta} = \frac{G_F^2 E_{\nu}^2}{4 \pi^2}\left[c_V^2(1+\cos\theta)S_V(q) + c_A^2(3-\cos\theta)S_A(q)  \right]
\eeq
where $E_{\nu}$ is the energy of the neutrino, $\cos\theta$ is the scattering angle and $q = 2 E_{\nu} \sin\theta/2$ is the momentum transfer to the medium in which the neutrino is scattering. The functions $S_V(q)$ and $S_A(q)$ are the vector and axial static structure factors, respectively, and they encode the properties of the medium. In the non-relativistic limit relevant here \footnote{The conditions of the neutrinosphere, $T\sim 5-10 \text{ MeV}$ and $n \sim 10^{-3}-10^{-1} n_0$ are such that the neutrons composing the medium can be approximated by a non-relativistic spin-1/2 field. In the non-relativistic limit, neutrinos couple to neutron density and spin.}, the structure factors are defined as Fourier transforms of equal time correlation functions
\begin{align*}
S_V(q) & = \int{d^3x~\e^{-i q\cdot x} \langle \delta n(x,0) \delta n(0,0) \rangle  }~, \\
S_A(q) & = \int{d^3x~\e^{-i q\cdot x} \langle \delta s_z(x,0) \delta s_z(0,0) \rangle  }~.
\end{align*}
where $n$ is the density of particles, and $s_z$ is the density of z component of spin. In a supernova, these structure factors are computed in a high temperature medium composed of primarily neutrons; we calculate here $S_V$ and $S_A$ in a hot medium composed of unitary fermions as a first step toward calculating the structure factors of the neutron gas. 

Various properties of the unitary gas have been computed in the past, with particular interest in the low temperature regime. Thermodynamic functions in the have been calculated by several groups~\cite{PhysRevA.88.063643,Jensen:2018opr,PhysRevA.78.023625,PhysRevA.85.051601,PhysRevLett.103.210403,PhysRevLett.110.090401}. These calculations are done in the vicinity of the superfluid phase transition and in the postulated ``pseudogap" regime, both of which occur at low temperature. The contact has also been calculated by several groups \cite{Jensen:2019zkr,PhysRevLett.106.205302,Goulko:2015rsa,PhysRevLett.121.130406}, and the vector static structure factor has been calculated by relatively fewer groups, with a zero temperature Quantum Monte Carlo calculation performed in \cite{PhysRevLett.110.055305}.

Due to the unique conditions of the neutrinosphere, we explore here a higher temperature regime than in previous studies; as a result we probe novel regions of parameter space of the unitary gas. Furthermore, we present here the first non-perturbative calculation of both the axial and vector structure factors of the unitary gas at the thermodynamic parameters found in the neutrinosphere. In particular, we calculate the structure factors of the unitary gas beyond the domain of validity of the commonly used virial expansion, which has been used to calculate structure factors perturbatively in the past~\cite{PhysRevC.98.015802} \footnote{The virial expansion has also been used to compute \emph{dynamic} properties of the unitary gas. For example the shear and bulk viscosities have very recently been computed in \cite{Hofmann:2019jcj}.}.

This paper is organized as follows: In Section \ref{theory} we describe the methods we used to numerically simulate the unitary gas. In Section~\ref{tuning} we describe how we tune lattice parameters of the hamiltonian to satisfy the unitary condition. In Section~\ref{observables} we derive the observables of interest. In Section~\ref{results} we show the results of our lattice calculation and in Section~\ref{conclusions} we state our conclusions.

%yoram's paper:\cite{Jensen:2018opr}

%Logic:
%we compute the structure factors of the unitary gas under supernova conditions
%
%why do that?
%
%there is a region around a protoneutron star called the neutrinosphere, that is at high temperature and density where the DoF are basically neutrons, and neutrino scattering is important in here. 20-30 percent corrections to scattering rates can make or break an explosion.
%
%to calculate scattering rate
\section{Formalism} \label{theory}

We will use the following lattice hamiltonian to describe the unitary gas:
\beq
\label{eq:ham-lat-unitary}
H -\mu N= \underbrace{  \sum_{\sigma; xy}{\psi^{\dagger}_{\sigma; x} k_{xy} \psi_{\sigma; y}}   }_{K}
\underbrace{ 
- \mu \sum_{\sigma; x}{\psi^{\dagger}_{\sigma;x }\psi_{\sigma; x}} - \frac{G}{\Delta x^3} \sum_{x}{\psi^{\dagger}_{1;x}\psi_{1;x}\psi^{\dagger}_{2;x}\psi_{2;x}}   }_{V}.
\eeq
Here $\psi_{\sigma;x}$ annihilates a fermion with spin $\sigma$ at site $x$ on a 3D spatial lattice, the fermionic fields are normalized such that $\{\psi_{\sigma;x},\psi^{\dagger}_{\sigma';x'}\}=\delta_{\sigma \sigma'}\delta_{x x'}$ and the matrix $k_{xy}$ is given by
\beq
k_{xy} = \frac{1}{2M\Delta x^2}\sum_{p}{p^2\e^{-ip\cdot (x-y)} }~,
\eeq 
where $p$ is a lattice momentum. The non-local hopping term generates the correct, continuum dispersion relation, instead of a lattice approximation that would be generated if a local hopping term had been used. The non-local structure of the hamiltonian will not cause any difficulty in the simulations as this term is treated in momentum space. Of course, the hamiltonian in~\eq{eq:ham-lat-unitary} is a discretized version of 
\beq
\label{eq:cont-hamiltonian}
H=\int{d^3x \Big[\psi^{\dagger}_{\sigma}\big(-\frac{\nabla^2}{2 M}-\mu\big)\psi_{\sigma}-\frac{G}{2}(\psi_{\sigma}^{\dagger}\psi_{\sigma})^2 \Big]}~,
\eeq 
which, for a properly renormalized $G$, describes non-relativistic fermions at unitarity. As we will show latter, if we regularize the
theory using the lattice discretization in~\eq{eq:ham-lat-unitary}, the unitarity point is given by $G\approx 5.14435 \Delta x/M$.

Our object of study is the partition function (and its derivatives with respect to local spacetime sources)
\beq\label{eq:trotter}
Z=\Tr(e^{-\beta(H-\mu N)}) = \Tr\left(\prod_{t=1}^{N_t}{e^{-\Delta t K}e^{-\Delta t V}}\right)+\mathcal{O}(\Delta t^2)~,
\eeq 
where $\Delta t = \beta/N_t$ \footnote{That this trotterization is correct to $\mathcal{O}(\Delta t^2)$ can be seen by noting that \newline $\Tr(\prod_{t=1}^{N_t}{e^{-\Delta t K}e^{-\Delta t V}}) = \Tr(\prod_{t=1}^{N_t}{e^{-\frac{\Delta t}{2} K}e^{-\Delta t V}}e^{-\frac{\Delta t}{2} K})$ by cyclicity of the trace.}. 
The numerical Monte Carlo method we will use is very similar to the one in \cite{PhysRevB.40.506,PhysRevD.24.2278}.
The four-fermion interaction can be written in terms of fermion bi-linears by noting that
\beq
e^{-\Delta t V}  = \prod_{x}{\exp\left({\Delta t \mu (n_{1x}+n_{2x})+\frac{\Delta t G}{\Delta x^3} n_{1x}n_{2x} }\right)}
= \prod_{x}\frac1{f_0}\int{ dA_x e^{-\frac{1}{\hat g}(\cosh A_x -1)} e^{(n_{1x}+n_{2x})(A_x +\hat \mu)} }~,
\eeq 
with $n_\sigma = \psi^\dagger_\sigma \psi_\sigma$, provided $\mu, G$ are related to $\hat\mu, \hat g$ through
$e^{\mu \Delta t}  = \frac{f_1}{f_0} e^{\hat \mu}$ and $e^{\frac{G \Delta t}{\Delta x^3}}={\frac{f_0 f_2}{f_1^2}}$, where the functions $f(\hat g)$ are
\beq
f_\alpha(\hat g) \equiv \int_{-\infty}^\infty dA \exp\left(-\frac{\cosh A-1}{\hat g}+\alpha A\right).
\eeq 
 Finally, using the identity
\beq
\Tr \left(e^{-\sum{\psi^{\dagger}_x (M_1)_{xy}\psi_y}} e^{-\sum{\psi^{\dagger}_x (M_2)_{xy}\psi_y}}\cdots e^{-\sum{\psi^{\dagger}_x (M_N)_{xy}\psi_y}}   \right) = \det\left(\mathbf{1}+e^{-M_N}\cdots e^{-M_1}\right)
\eeq
for fermionic operators $\psi^{\dagger}_x,\psi_y$, we see that
\beq\label{eq:partition-function-sausage}
Z \approx \Tr\left(\prod_{t=1}^{N_t}{e^{-\Delta t K}e^{-\Delta t V}}\right) =  f_0^{-N_s N_t}\int{DA\,e^{-S_g(A)}
\det\left[\mathbf{1}+e^{N_t\hat\mu} B^{-1}C(A_{N_t-1})\cdots B^{-1}C(A_{0}) \right]^2} 
\eeq where
\beq
B_{xx'} = \sum_{p}{e^{-ip\cdot(x-x')}e^{\hat{\gamma}\frac{p^2}{2}}} , 
\ \ \
C(A_t)_{x x'} = \delta_{x x'} e^{A_{x,t}},
\ \ \ 
\hat \gamma\equiv\frac{\Delta t}{M\Delta x^2 } ,
\eeq 
$N_s$ is the number of spatial sites, and $DA = \prod_{xt}dA_{xt}$.

The same partition function, up to a constant multiplicative factor, 
can also be derived from the path integral over the Euclidean lattice action:
\bea
S &=& \frac{1}{\hat g}\sum_{x,t}{(\cosh(A_{x,t})-1)}
+
\sum_{\sigma; x,y,t}{\hat\psi^{\dagger}_{\sigma;x,t}B_{xy} \hat\psi_{\sigma;y,t}}- \sum_{\sigma;x,t}{\hat \psi^{\dagger}_{\sigma;x,t+1}\text{e}^{A_{x,t}+ \hat \mu}\hat \psi_{\sigma;x,t}} \nn\\
&=&
\frac{1}{\hat g}\sum_{x,t}{(\cosh(A_{x,t})-1)}
+
\sum_{\sigma; xtx't'}{\hat\psi^{\dagger}_{\sigma;x,t}D_{xtx't'}\hat\psi_{\sigma;x',t'}},
\label{eq:our-action}
\eea 
where we define the fermion matrix $D_{xt,x't'} = B_{xx'} \delta_{tt'} - e^{A_{xt}+\hat\mu} \delta_{t+1t'}$ (we implement anti-periodic boundary conditions in time by taking $\delta_{N_t,N_{t-1}}  \equiv -\delta_{0,N_{t-1}}$.)
Indeed, the integration over Grassmann variables $\hat\psi$ leads to
\beq
\label{eq:partition-function}
\int{DA D\hat\psi D\hat\psi^{\dagger}~e^{-S[A,\psi^{\dagger},\psi]}} = \int{DA~e^{-S_g[A]}\det( D)^2}~,
\eeq 
where $S_g(A) \equiv \frac{1}{\hat g} \sum_{x,t}{(\cosh(A_{x,t})-1)}$ and $D$ is the fermion matrix. The determinant $\det D$ can be written, 
by repeated use of the Schur complement identity,
%\beq
%\det D = \det d \det (a-b d^{-1} c),\quad\text{where}\quad
%D=
%\begin{pmatrix}
%a & b\\
%c & d\\
%\end{pmatrix},
%\eeq 
as
\beq
\det D = \det B^{N_t} \det\left[\mathbf{1} + e^{N_t\hat\mu}
B^{-1}C(A_{N_t-1})...B^{-1}C(A_{0}) \right] = \det(\mathbf{1} + U) \,.
\label{eq:hubb}
\eeq 
where $U\equiv e^{N_t\hat\mu}
B^{-1}C(A_{N_t-1})...B^{-1}C(A_{0})$. Therefore, the expression for the derived from the action ~\eq{eq:partition-function} is equivalent to the one derived from the Hamiltonian in~\eq{eq:partition-function-sausage}.

A few words about the Monte Carlo calculation of expression in \eq{eq:partition-function-sausage} (or, equivalently, of expression \eq{eq:partition-function}) are in order. Firstly, since $D$ is a real matrix $\det D$ is real, $\det D^2 \geq 0$ and the theory is sign-problem free.  Secondly, the matrix $B^{-1}$ is independent of $A_{xt}$ and can be computed once at the beginning of the calculation and used thereafter. The matrices $C(A_t)_{xx'}$, which depend on $A_{xt}$, are diagonal and their multiplication has a computational cost of order ${\cal O}(N_s^2)$, where $N_s$ is the number of spatial sites (instead of the ${\cal O}(N_s^3)$ cost for dense matrices). Finally, we use the Hybrid Monte Carlo algorithm of~\cite{PhysRevD.98.054514} to sample fields.

\section{Unitarity Condition}
\label{tuning}

To obtain the unitary gas in the continuum and infinite volume limits, we tune the parameters of the lattice theory such that there is a zero energy two-particle bound state in the ${}^1 S_0$ scattering channel at $\mu =0$. The binding energy of the two-particle state can be determined from the position of the appropriate pole in the scattering amplitude. For this model, conservation of particle number allows us compute exactly the scattering amplitude by summing over all Feynman diagrams that contribute to two-particle scattering processes.

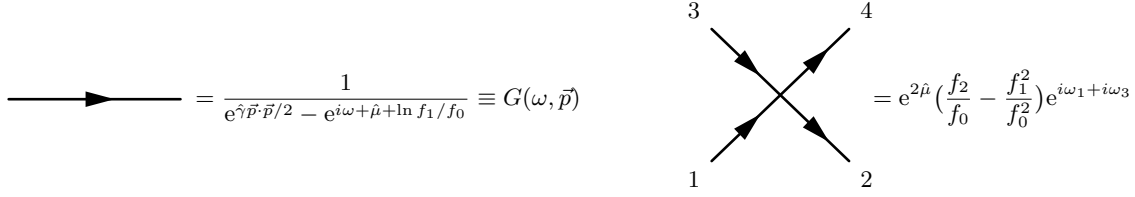
\begin{figure}[t!]
\begin{subfigure}{.5\textwidth}
\centering
\begin{fmffile}{propogator}
\begin{equation}
\vcenter{\hbox{
%\begin{gathered}
\begin{fmfgraph*}(65,50)
\fmfleft{i1}
\fmfright{o1}
\fmf{fermion}{i1,o1}
\end{fmfgraph*}
%\end{gathered} 
}}
=\frac{1}{\e^{\hat \gamma \vec{p}\cdot\vec{p}/2}-\e^{i\omega+\hat \mu + \ln f_1/f_0}}  \equiv G(\omega,\vec{p}) \nonumber
\end{equation}
\end{fmffile}
%\caption{Fermion propagator in the lattice theory}
\label{fig:sfig1}  
\end{subfigure}%
\begin{subfigure}{.5\textwidth}
  \centering
  
\begin{fmffile}{vertex} % This is a bad sample.
\begin{equation}
\begin{gathered}
\begin{fmfgraph*}(65,50)
\fmfleft{i1,i2}

\fmfright{o1,o2}

%\fmflabel{$(\omega_1,{\bf p}_1)$}{i1}
\fmflabel{$1$}{i1}
\fmflabel{$2$}{o1}
\fmflabel{$3$}{i2}
\fmflabel{$4$}{o2}

\fmf{fermion}{i1,v,o1}
\fmf{fermion}{i2,v,o2}
%\fmflabel{$a$}{i1}
\end{fmfgraph*}
\end{gathered}
=\e^{2\hat\mu}\big(\frac{f_2}{f_0}-\frac{f_1^2}{f_0^2}\big)\e^{i\omega_1+i\omega_3} \nonumber
\end{equation}
\end{fmffile}
%  \caption{Vertex in the lattice theory}
  \label{fig:sfig2}
\end{subfigure}
\caption{Feynman rules corresponding to the action in~\eq{eq:inf-vol-lat-act} for propagator and vertex.}
\label{fig:feynman-rules}
\end{figure}

To compute the Feynman rules we start from the action \eq{eq:partition-function}, and integrate over the auxiliary fileds:
%The two particle binding energy can be computed exactly on the lattice. To do so, it is first necessary to explicitly integrate out the auxiliary fields in~\eq{eq:partition-function}. This can be done by expanding out the exponential of Grassmann variables then integrating over auxiliary fields:
\beqs
&\int_{-\infty}^\infty dA \exp\left(-\frac{\cosh A-1}{\hat g}\right) \, \exp[- \hat\psi^\dagger e^{A+\hat\mu} \hat\psi]
=\int_{-\infty}^\infty dA \exp\left(-\frac{\cosh A-1}{\hat g}\right)\, [ 1 - \hat\psi^\dagger e^{A+\hat\mu} \hat\psi
+\frac{1}{2} (\hat\psi^\dagger e^{A+\hat\mu} \hat\psi)^2 ] \\
&\quad= f_0 - f_1 \hat\psi^\dagger e^{\hat\mu} \hat\psi +\frac{1}{2} f_2 (\hat\psi^\dagger e^{\hat\mu} \hat\psi)^2
= f_0 \exp \left[\frac{f_1}{f_0}\hat\psi^\dagger e^{\hat\mu} \hat\psi + \left(\frac{f_2}{f_0}-\frac{f_1^2}{f_0^2} \right) \frac{1}{2}(\hat\psi^\dagger e^{\hat\mu} \hat\psi)^2 \right] .
\eeqs
% One therefore has
%\beq
%Z =\text{const}\times \int{D\psi D\psi^{\dagger}e^{-S_f[\psi,\psi^{\dagger}]}}~,
%\eeq
%where
The action in terms of Grassmann variables is then:
\beq
S_f = \sum_{\sigma; x,y,t}{\psi^{\dagger}_{\sigma;x,t}B_{xy}\psi_{\sigma;y,t}} - e^{\hat \mu}\frac{f_1}{f_0} \sum_{\sigma;x,t}{\psi^{\dagger}_{\sigma;x,t+1}\psi_{\sigma;x,t}}-e^{2\hat\mu}\left(\frac{f_2}{f_0}-\frac{f_1^2}{f_0^2}\right)\sum_{x,t}{\psi^{\dagger}_{1;x,t+1}\psi_{1;x,t}\psi^{\dagger}_{2;x,t+1}\psi_{2;x,t} }\,.
\eeq
%Taking a lattice of infinite extend and converting to momentum space one obtains
For an infinite lattice in both space and time, the momentum space action reads
\begin{align}
\label{eq:inf-vol-lat-act}
S_f = & \int_{-\pi}^{\pi}{d^4 p ~\psi^{\dagger}_{\sigma}(\omega,p)\Big[e^{\hat \gamma \vec{p}\cdot\vec{p}/2}-e^{i\omega+\hat \mu + \ln f_1/f_0}\Big]\psi_{\sigma}(\omega,p)} \\ \nonumber
 & -e^{2\hat\mu}\left(\frac{f_2}{f_0}-\frac{f_1^2}{f_0^2}\right)\int_{-\pi}^{\pi}{d^4 p_1 d^4 p_2 d^4 p_3 d^4 p_4~\delta^4(p_1+p_3-p_2-p_4)e^{i\omega_1 + i \omega_3}\psi^{\dagger}_1(p_1)\psi_1(p_2)\psi^{\dagger}_2(p_3)\psi_2(p_4)}\, ,
\end{align}
leading to the Feynman rules in~\fig{fig:feynman-rules}.
%The Feynman rules corresponding to this action are given in \fig{fig:feynman-rules}. 
%The binding energy of the two-particle state can be extracted from a pole in the scattering amplitude. For this theory, the scattering amplitude can be computed exactly. 

 %feynman rules figure

In terms of diagrams, the scattering amplitude is given by the  sum of graphs shown in~\fig{fig:bubble-sum}. Choosing kinematics $(E/2,\vec{p})$ and $(E/2,-\vec{p})$ on the external legs, the scattering amplitude has the analytic expression
\beq
\label{eq:scat-amp}
-\mathcal{A} = \frac{1}{\Big[e^{2\hat\mu}\big(\frac{f_2}{f_0}-\frac{f_1^2}{f_0^2}\big)e^{iE}\Big]^{-1}-L(E)}~,
\eeq
where 
\beq
L(E)  = \int_{-\pi}^{\pi}{\frac{dq_0}{2\pi}}\int_{-\pi}^{\pi}{\frac{d^3 q}{(2\pi)^3}G(q_0+E/2,\vec{q})G(-q_0+E/2,-\vec{q})}
\eeq
is the fermion-fermion loop iterated in the bubble sum. To work at $\mu=0$, using the map derived above, we set $\hat\mu = \log\tfrac{f_0}{f_1}$ and the unitarity condition, that the amplitude has a pole at $E=0$ at zero density, becomes:
%Since we require a zero-energy bound state at $\mu=0$, using the map between physical and lattice quantities $\mu \Delta t = \hat \mu + \log(f_1/f_0)$, we can set $e^{\hat \mu} = \frac{f_0}{f_1}$ in \eq{eq:scat-amp}, and therefore the zero-energy bound state condition is
\beq
\label{eq:unitary-cond}
\frac{f_2 f_0}{f_1^2}-1= \frac{1}{L(0)}\,,\quad\text{with}\quad L(0) = \int_{-\pi}^{\pi}{\frac{d^3 q}{(2\pi)^3} \frac{1}{e^{\hat \gamma q^2}-1} } \,.
\eeq
Given any $\hat \gamma = \Delta t/(M \Delta x^2)$, this condition fixes the lattice coupling $\hat g$. In principle we
could enforce this condition for any value of $\Delta x$ and $\Delta t$ and then choose any sensible path to take the limit 
$\Delta t\to 0$, $\Delta x\to0$. In practice we find it advantageous to fix $\Delta x$, fix $G$ from the unitarity condition:
\beq
\frac{MG}{\Delta x} = \lim_{\hat\gamma\to0} \frac1{\hat\gamma L(0)} = 5.14435\ldots \,,
\eeq
and take the limit $\Delta t\to 0$ while $\hat g$ is fixed by $\frac{\Delta t G}{\Delta x^3} = \log(\frac{f_2 f_0}{f_1^2})$.
This has the advantage that for fixed $\Delta x$, the discretized partition function has ${\cal O}(\Delta t^2)$ errors, since
for this parameter choice the partition function derived from the action is the same as the approximation introduced in~\eq{eq:trotter}.
This is the condition we use to fix couplings in Monte Carlo calculations. Of course, a subsequent spatial continuum limit $\Delta x
\to0$ is required to get the final results.
 
%Choosing to take the continuum limit by tuning  $\Delta t \rightarrow 0$  at fixed $\Delta x$ makes the trotterization ({eq:trotter}) exact and, according to our experience, leads to a smoother continuum limit. For that reason, we will define the unitarity condition in the $\Delta t \rightarrow 0$ limit or, equivalently, in the $\hat \gamma\rightarrow 0$ limit.
%Using the mapping $\frac{\Delta t G}{\Delta x^3} = \log(\frac{f_2 f_0}{f_1^2})$, and using the fact that 
%\beq
%L(0) = \int_{-\pi}^{\pi}{\frac{d^3 q}{(2\pi)^3} \frac{1}{e^{\hat \gamma q^2}-1} } \underset{\hat \gamma \rightarrow 0}{\longrightarrow} \frac{0.19488...}{\hat \gamma}
%\eeq
%$L(0) \underset{\hat \gamma \rightarrow 0}{\longrightarrow} \frac{0.19488...}{\hat \gamma}$ we have
%\beq
%\frac{MG}{\Delta x} = 5.14435...
%\eeq
%which is the unitary condition~\eq{eq:unitary-cond} in the $\hat \gamma\rightarrow 0$ limit. This is the condition we use to fix couplings in Monte Carlo calculations. 
\begin{figure}[t!]
\centering
\begin{fmffile}{bubble-sum}
\begin{equation}
-\mathcal{A} \quad= 
\parbox{30mm}
{
\begin{fmfgraph*}(100,70)
\fmfleft{i1,i2}
\fmfright{o1,o2}
\fmf{fermion}{i1,v,o1}
\fmf{fermion}{i2,v,o2}
%\fmflabel{$a$}{i1}
\end{fmfgraph*}
}
\quad+\quad
\parbox{30mm}
{
\begin{fmfgraph*}(100,70)
\fmfleft{i1,i2}
\fmfright{o1,o2}
\fmf{fermion}{i1,v}
\fmf{fermion}{i2,v}
\fmf{fermion,left=1}{v,v2}
\fmf{fermion,right=1}{v,v2}
\fmf{fermion}{v2,o1}
\fmf{fermion}{v2,o2}
\end{fmfgraph*}
}
\quad+\quad
\parbox{30mm}
{
\begin{fmfgraph*}(100,70)
\fmfleft{i1,i2}
\fmfright{o1,o2}
\fmf{fermion}{i1,v}
\fmf{fermion}{i2,v}
\fmf{fermion,left=1}{v,v2}
\fmf{fermion,right=1}{v,v2}
\fmf{fermion,left=1}{v2,v3}
\fmf{fermion,right=1}{v2,v3}
\fmf{fermion}{v3,o1}
\fmf{fermion}{v3,o2}
\end{fmfgraph*}
}
\quad+\quad ... 
\end{equation}
\end{fmffile}
\caption{The scattering amplitude $\mathcal{A}$ is given by the bubble sum above. To tune to unitarity on the lattice, we compute the bubble sum with lattice vertices and propagators on a lattice of infinite spacetime extent, corresponding to the zero temperature and infinite volume limits, and demand that there exist a zero energy bound state. }
\label{fig:bubble-sum}  
\end{figure} %bubble sum figure

\section{Observables}\label{observables}
We compute three observables: the density, the vector structure factor and the axial structure factor. The density is defined through the standard thermodynamic relation, and has the path integral expression
\beqs
\langle n \rangle & = \frac{1}{V}\frac{\partial}{\partial \mu}(T \log Z) \\ 
& = \frac{1}{\Delta x^3}\frac{1}{N_s}\frac{1}{Z}\int{DA~\e^{-S_g(A)}\det(\mathbf{1}+U(A))^2\Tr\Big[\mathbf{1}-(\mathbf{1}+U(A))^{-1}\Big] } \\
& = \frac{1}{\Delta x^3}\frac{1}{N_s} \langle \Tr\Big[\mathbf{1}-(\mathbf{1}+U)^{-1}\Big] \rangle\,,
\eeqs
%where the average is taken with respect to the probability density
%\beq
%p(A) = \frac{1}{Z}\e^{-S_g(A)}\det(\mathbf{1}+U(A))^2~.
%\eeq
where $U(A) = e^{N_t\hat\mu} B^{-1}C(A_{N_t-1})...B^{-1}C(A_{0})$. Our results are derived using cubic boxes with $N_x$ sites
in each dimension, so the total number of sites in a time slice is $N_s = N_x^3$.
The vector and axial structure factors are defined as Fourier transforms of equal time correlation functions of the density $n= n_1+n_2$ and spin $s_z= n_1-n_2$, respectively:
\beqs
S_V(q) &= \sum_x{\Delta x^3~\e^{-iq\cdot x}\langle \delta n(x,0) \delta n(0,0) \rangle }\,, \\ 
S_A(q) &= \sum_x{\Delta x^3~\e^{-iq\cdot x}\langle \delta s_z(x,0) \delta s_z(0,0) \rangle }\,.
\eeqs
Here we have defined the Heisenberg operators
\beqs
\delta n(x,t) & = n(x,t) - \langle n \rangle\,, \\ 
\delta s_z(x,t) & = s_z(x,t) - \langle s_z \rangle\,. \\ 
\eeqs
It is possible to derive path integral expressions of $S_V(q)$ and $S_A(q)$ by taking derivatives with respect to local sources. Coupling the theory to a spacetime dependent source on the $t=0$ timeslice, then setting the source to a constant after taking derivatives, one derives the following path integral expressions:
\beqs
\Delta x^6 \langle n_1(x,0) n_1(0,0) \rangle & =\frac{1}{Z}\int{DA~\e^{-S_g(A)}\det(\mathbf{1}+U(A))^2\left[\left(\mathbf{1}-(\mathbf{1}+U(A))^{-1}\right)_{00}\left(\mathbf{1}-(\mathbf{1}+U(A))^{-1}\right)_{xx}   \right] } \\ 
& + \frac{1}{Z}\int{DA~\e^{-S_g(A)}\det(\mathbf{1}+U(A))^2\left[ \left(\mathbf{1}-(\mathbf{1}+U(A))^{-1}\right)_{0x} \left(\mathbf{1}+U(A)\right)^{-1}_{x0}\right] } \\ 
\\
\Delta x^6 \langle n_1(x,0) n_2(0,0) \rangle & =\frac{1}{Z}\int{DA~\e^{-S_g(A)}\det(\mathbf{1}+U(A))^2\left[\left(\mathbf{1}-(\mathbf{1}+U(A))^{-1}\right)_{00}\left(\mathbf{1}-(\mathbf{1}+U(A))^{-1}\right)_{xx}   \right] }\,. \\ 
\eeqs
Since the action is invariant under $SU(2)$ spin rotations on the lattice (and in the continuum) $\langle n_1(x,0) n_1(0,0) \rangle = \langle n_2(x,0) n_2(0,0) \rangle$ and
$ \langle n_1(x,0) n_2(0,0) \rangle = \langle n_2(x,0) n_1(0,0) \rangle $, therefore the two correlation functions above are the only two independent density-density correlation functions. We compute these in our lattice calculations, then form the necessary linear combinations to compute $S_V$ and $S_A$. We also take advantage of translation invariance and hypercubic rotation invariance by summing over all such transformations in our calculation. This process reduces stochastic fluctuations of observables.

%below is some text on susceptibilities

%We also compute the both the spin and density susceptibilities. The spin susceptibility is given by
%\begin{align}
%\frac{\langle S^2 \rangle-\langle S \rangle^2}{V} & = T \Big(\frac{\partial s}{\partial h} \Big) \\
%& = \frac{T^2}{V} \frac{\partial^2}{\partial h^2} \text{ln}Z \\
%& =\frac{2}{\Delta x^3 N_x^3} \langle \Tr \big((\mathbf{1}+U)^{-1}\big)-\Tr \big((\mathbf{1}+U)^{-2}\big) \rangle~,
%\end{align}
%while the density susceptibility is 
%\begin{align}

\section{Results} \label{results}
Since the interactions in a unitary gas do not have any scale (the scattering length is infinite) the static structure factors $S_{V,A}$ of the unitary gas are function of the momentum $q$, chemical potential $\mu$, temperature $T$ and mass $M$. The dimensionless quantity $S_{V,A}/n$ (where $n$ is the density of the system)  is therefore a dimensionless function of only two variables
\begin{align}
S_V(q)/\langle n \rangle & = F_V\left(\frac{q}{q_{T}},z\right) \\ \nonumber
S_A(q)/\langle n \rangle & = F_A\left(\frac{q}{q_{T}},z\right)~,
\end{align} with $q_T=\sqrt{6MT}$ and $z=e^{\mu/T}$. The presence of a lattice introduces three unphysical scales, the finite volume and the lattice spacings $\Delta t$ and $\Delta x$. However, we will demonstrate below that our calculations are performed at large enough volumes and small enough $\Delta t, \Delta x$  that the dependence of $F_{V,A}$ on $q_T \Delta x$ and $T\Delta t$ can be neglected. 

Since the supernova collapse simulations are an important motivation for our calculations, we will concentrate on a subset of parameters relevant for the neutrino propagation on the neutrinosphere, namely, densities in the range of $0.001 n_0 < n < 0.03 n_0$ (where $n_0$ is the density of nuclear matter) and $3 \MeV < T < 10 \MeV$, corresponding to fugacities in the  $ 0.01 \alt z \alt 3$ range. This range of values is motivational only. A realistic evaluation of structure factors to be used in supernova modeling should include a more realistic neutron-neutron interaction (with  a non-zero  scattering length and effective range) as well as a small amount of protons and light nuclei. This corrections will be investigated in a  future study. 

\begin{figure}[t!]
\includegraphics[width=0.32\textwidth]{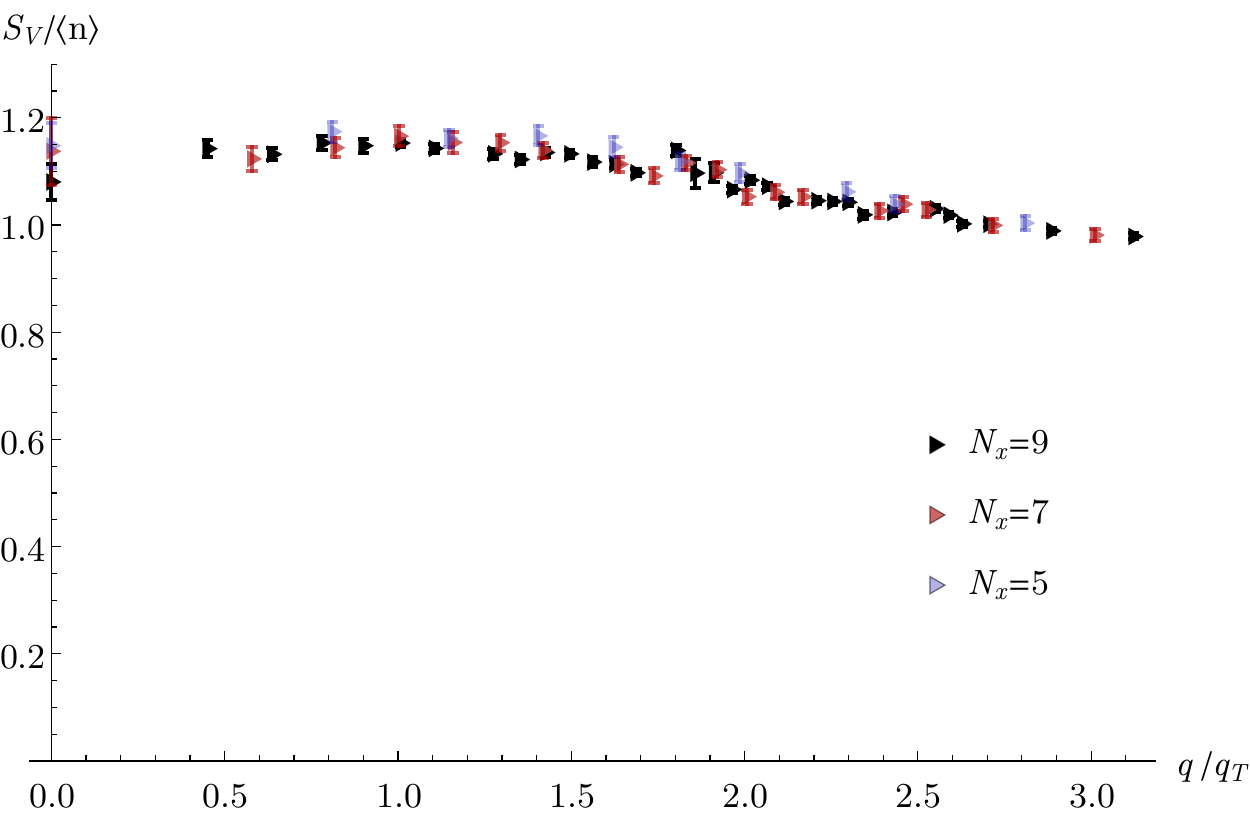}
\includegraphics[width=0.32\textwidth]{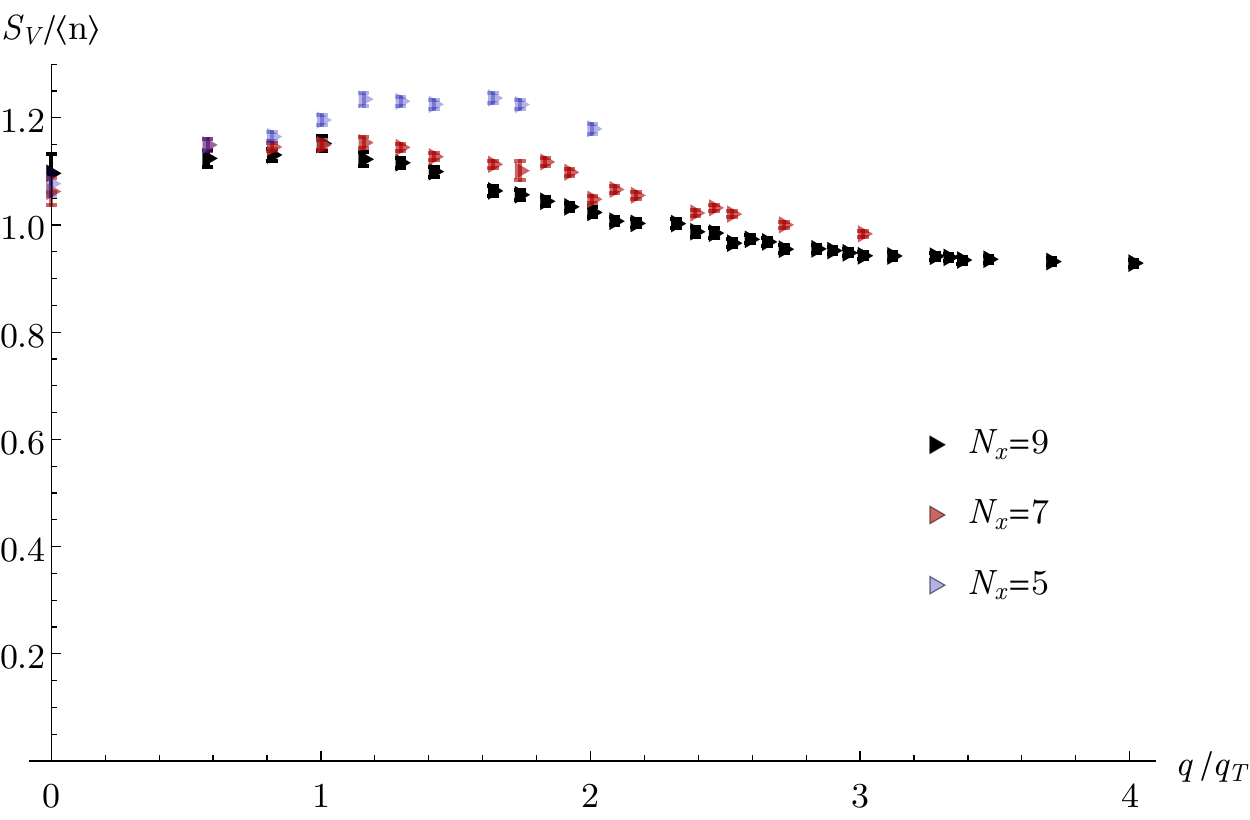}
\includegraphics[width=0.32\textwidth]{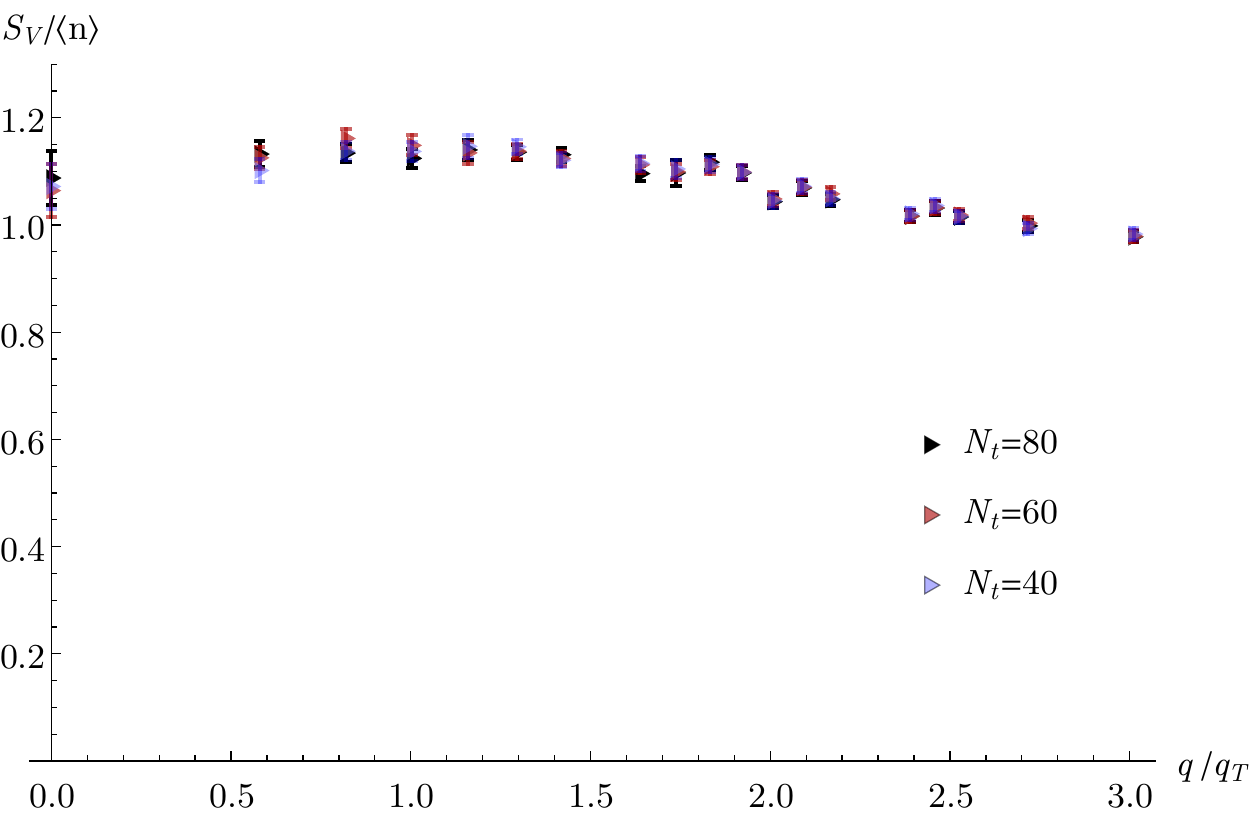} \\
\includegraphics[width=0.32\textwidth]{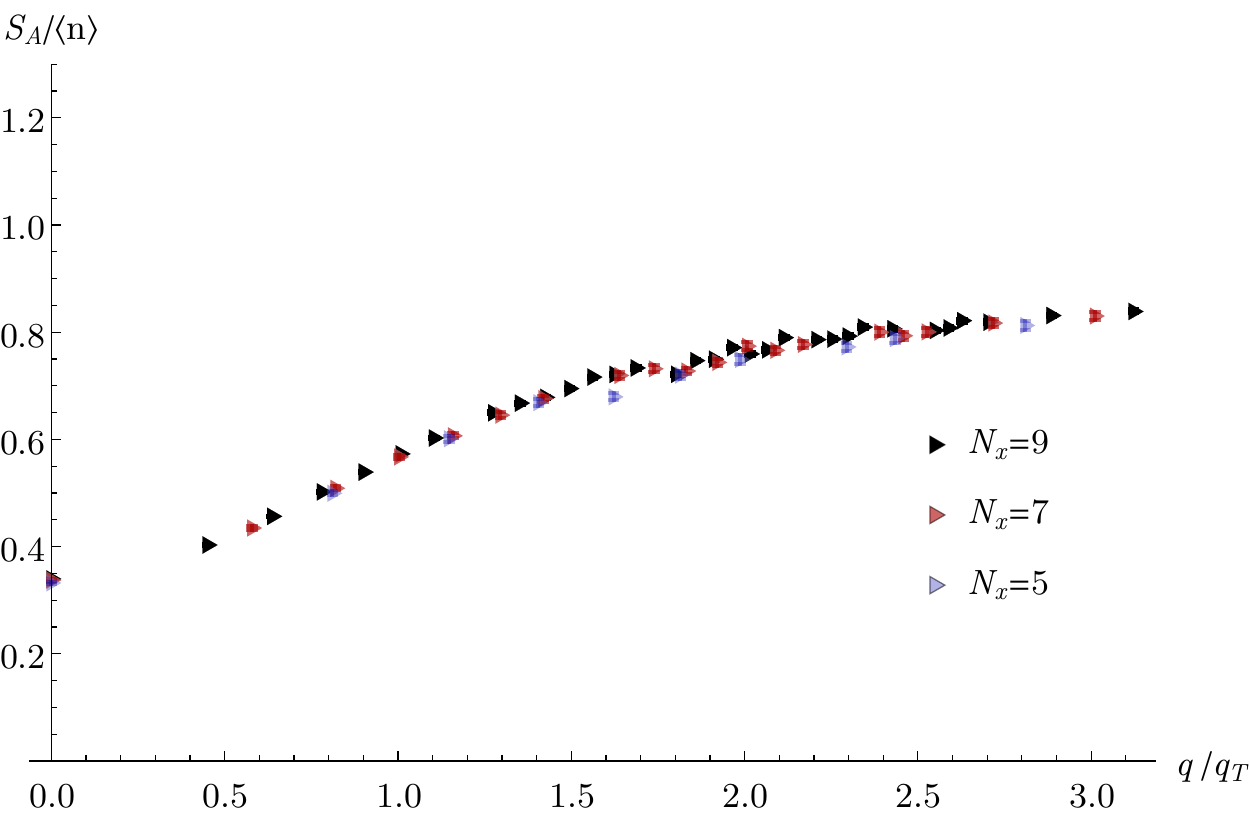}
\includegraphics[width=0.32\textwidth]{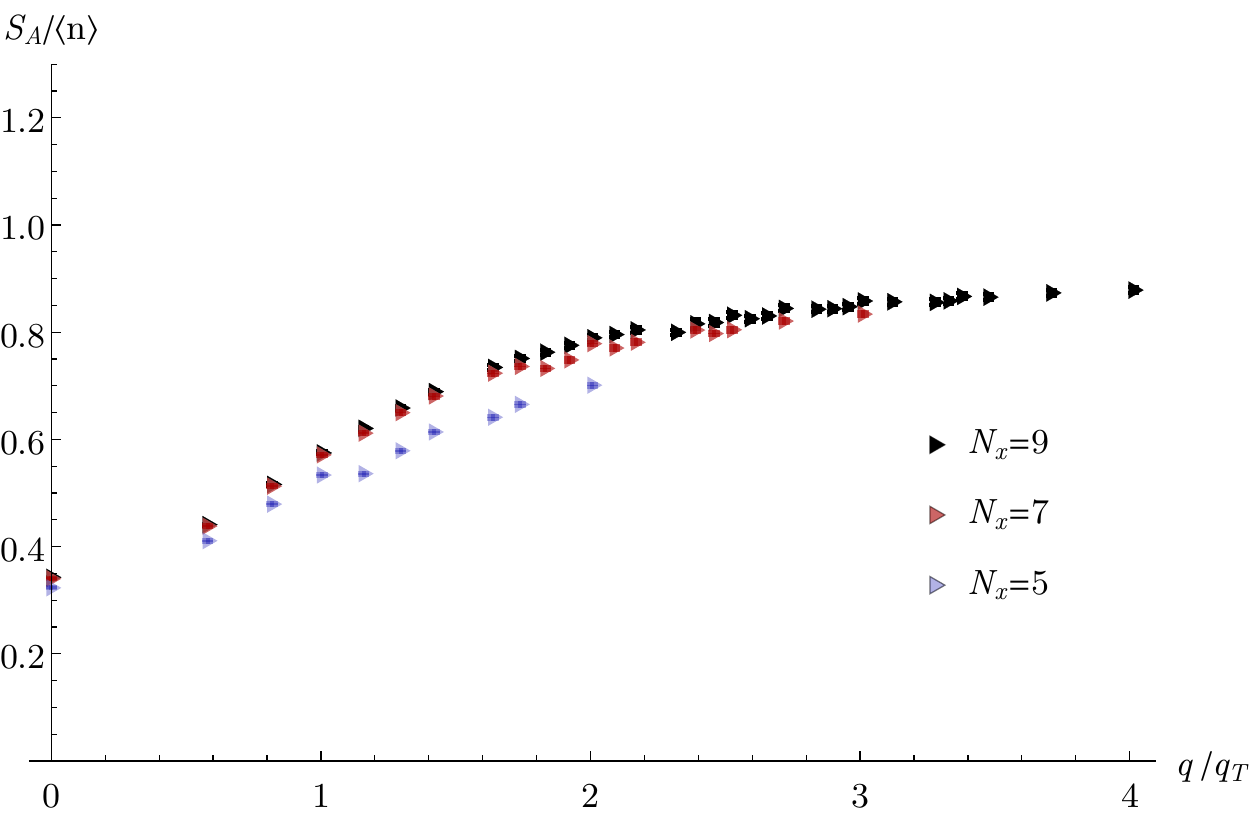}
\includegraphics[width=0.32\textwidth]{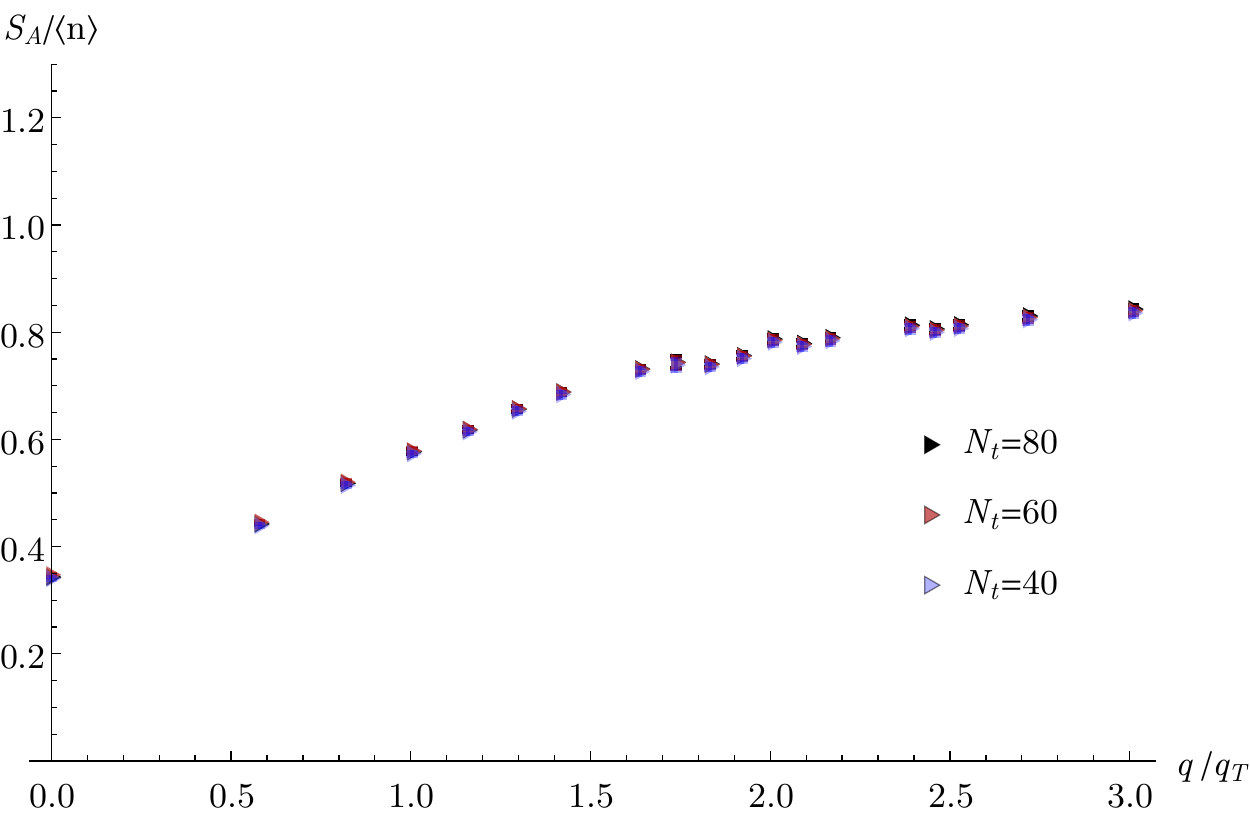}
\caption{Extrapolations of the vector (top row) and axial (bottom row) structure factors at $z=1.0$. The first column are infinte volume limits, the second column is the $\Delta x \rightarrow 0$ limit and the last column is the $\Delta t \rightarrow 0$ limit. The structure factors vary little as the infinite volume and time continuum limit are taken. There is more variation of the structure factors as the spatial continuum limit is taken, however the the structure factors appear more or less converged on the finest lattice spacing.}
\label{fig:z1p0extrap}
\end{figure}

\begin{figure}[t!]
\includegraphics[width=0.32\textwidth]{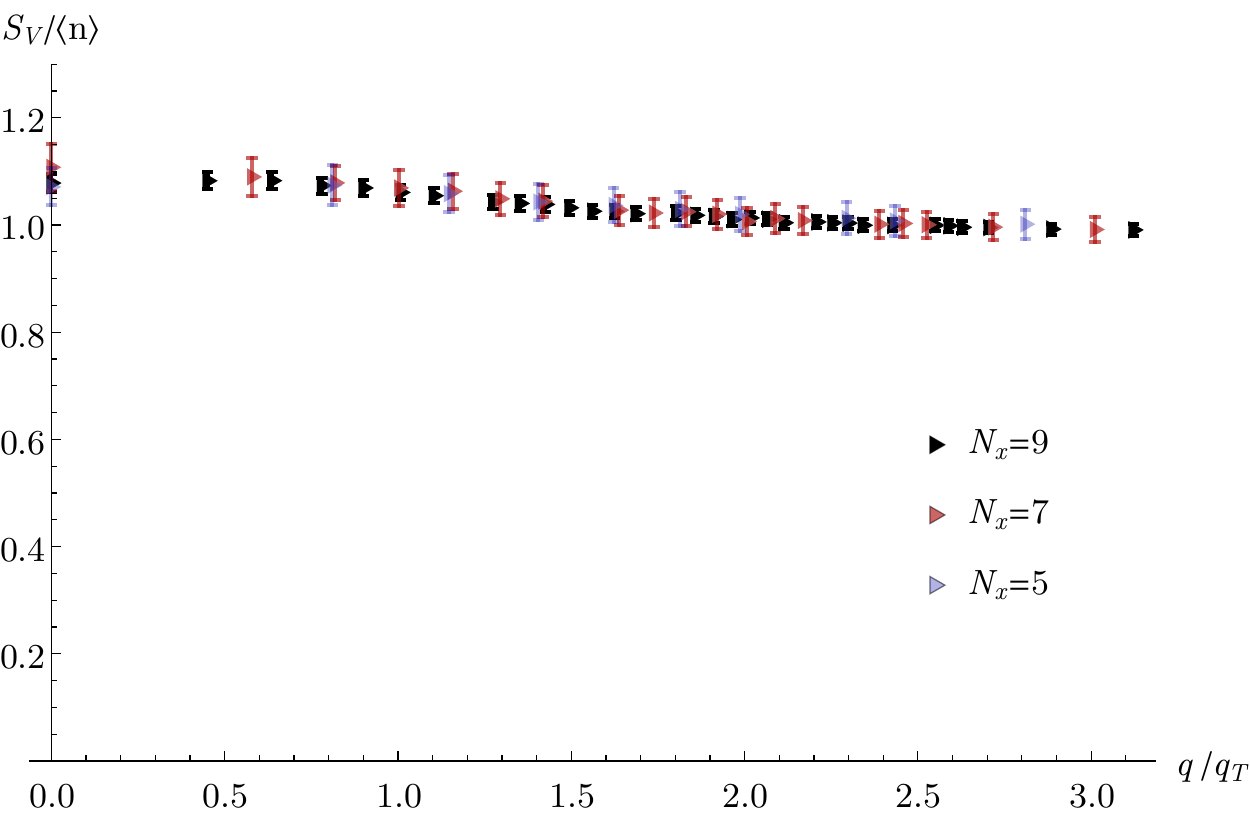}
\includegraphics[width=0.32\textwidth]{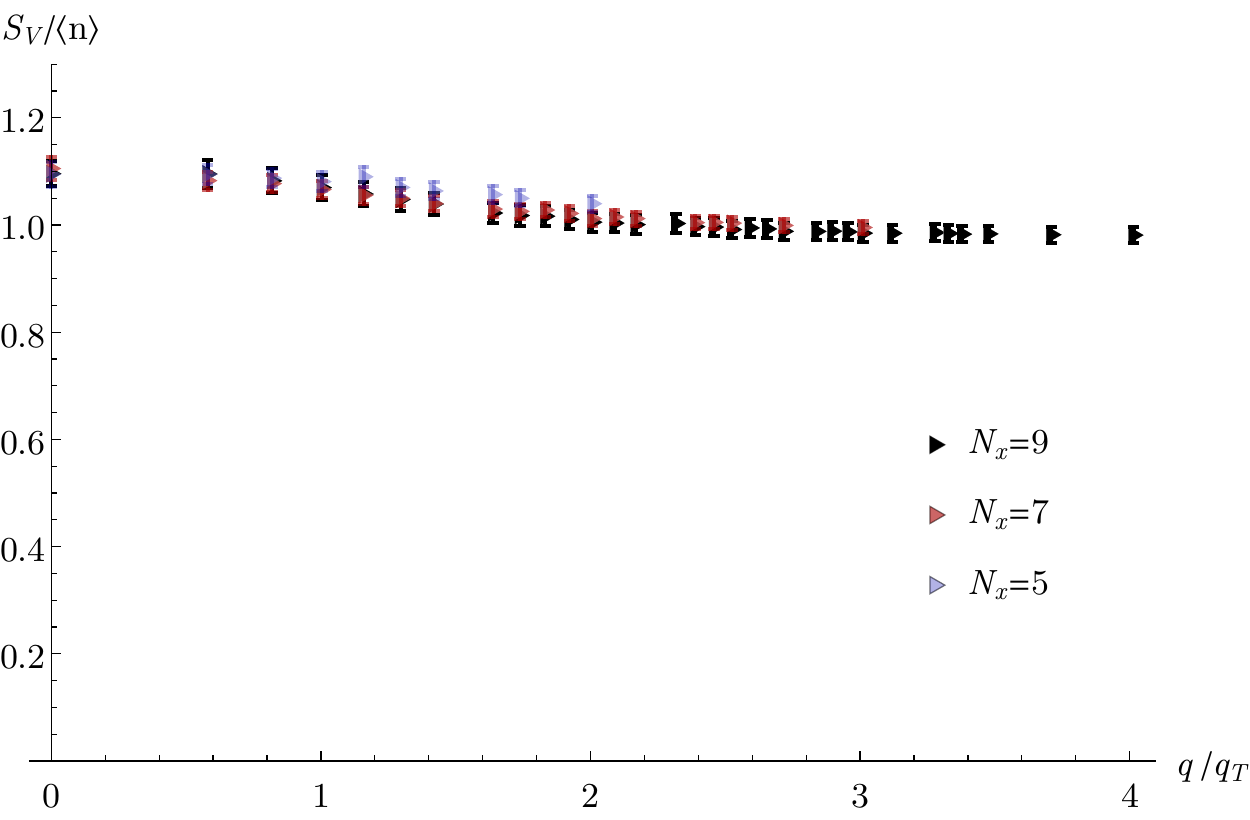}
\includegraphics[width=0.32\textwidth]{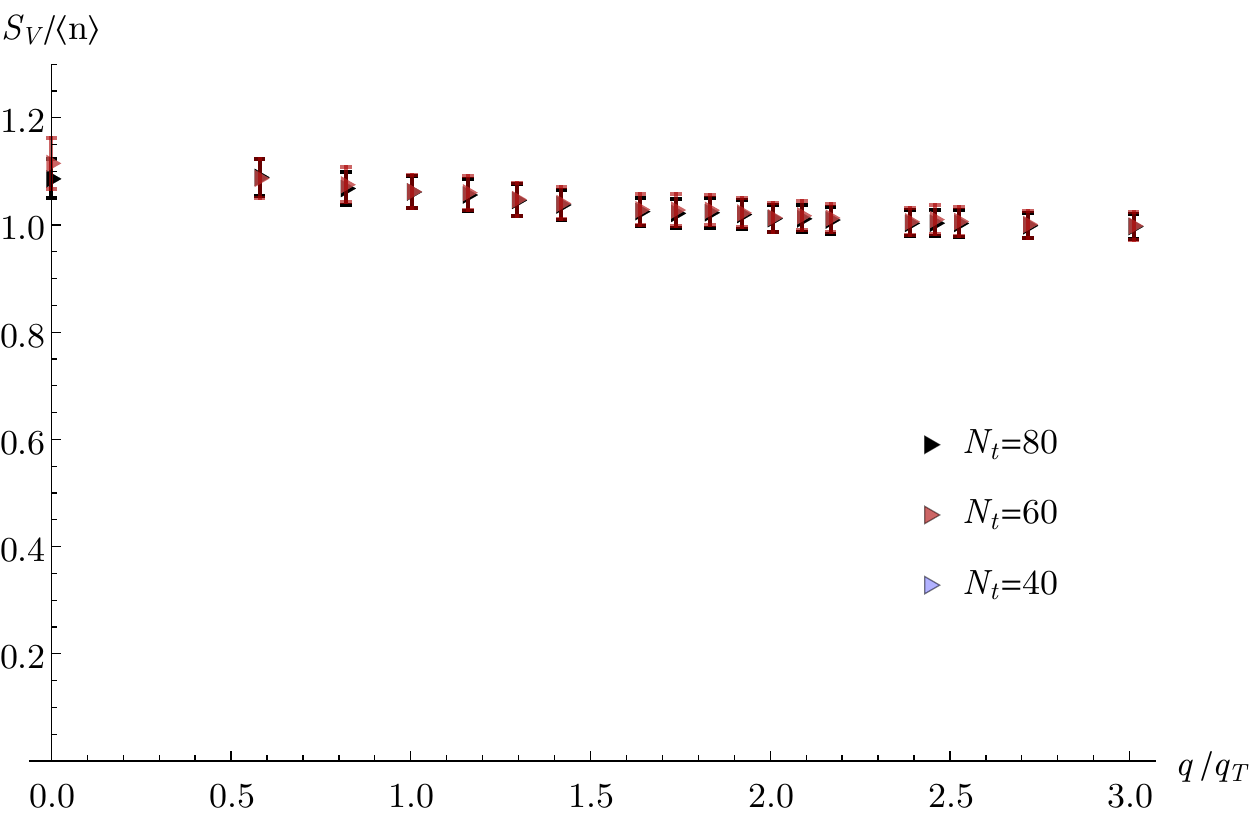} \\
\includegraphics[width=0.32\textwidth]{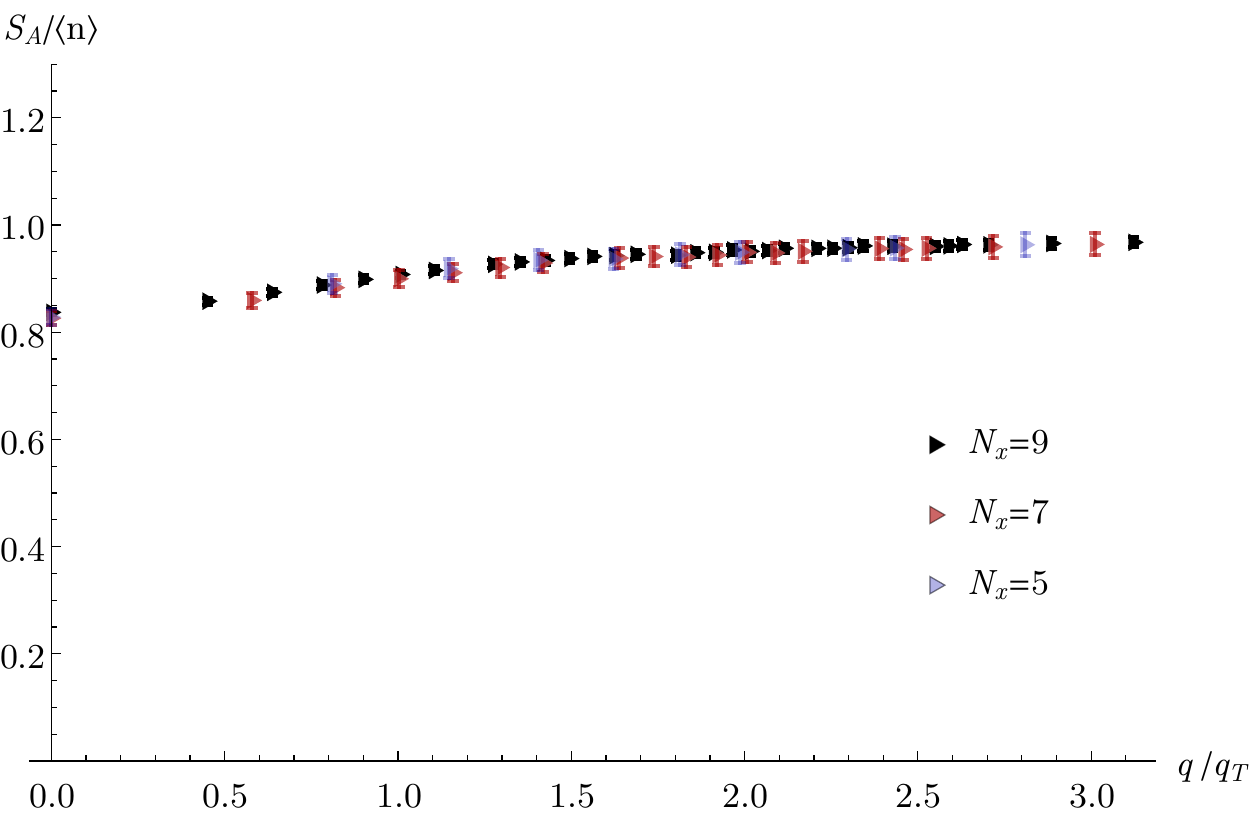}
\includegraphics[width=0.32\textwidth]{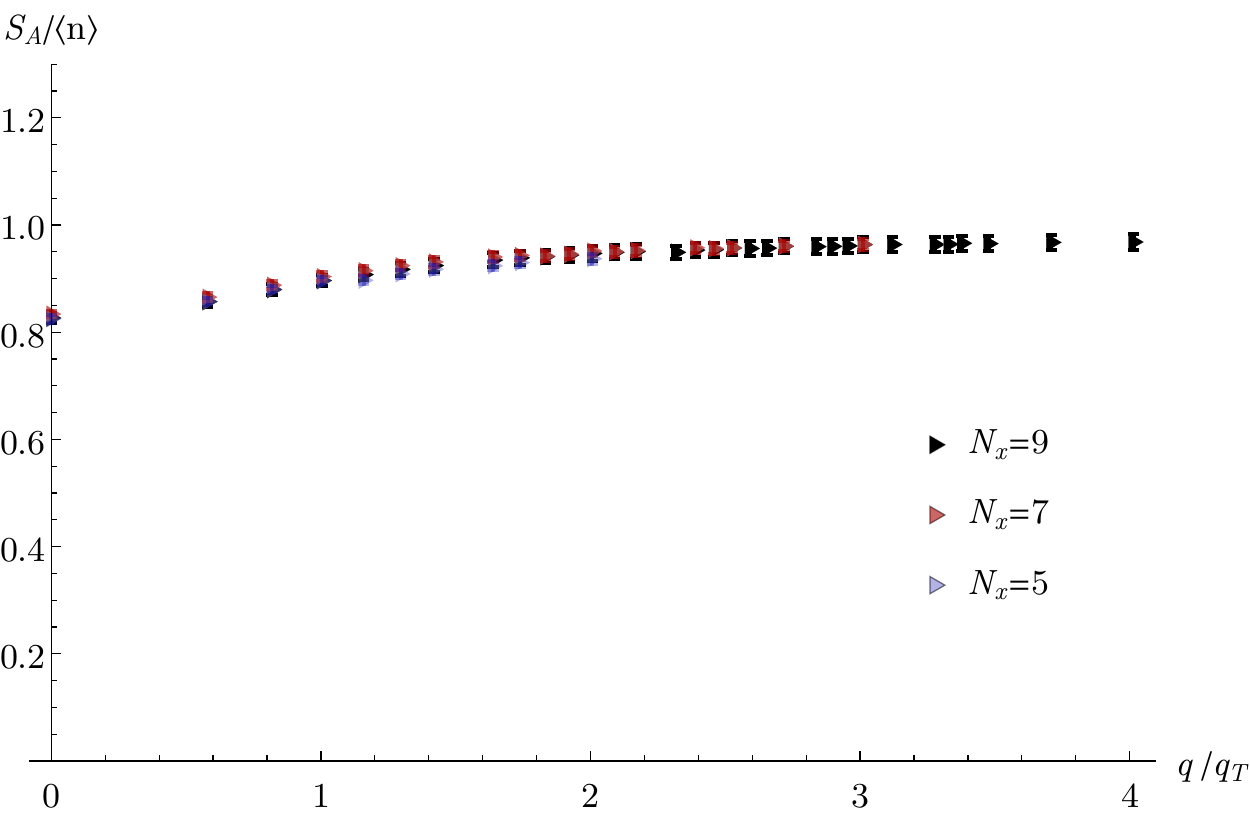}
\includegraphics[width=0.32\textwidth]{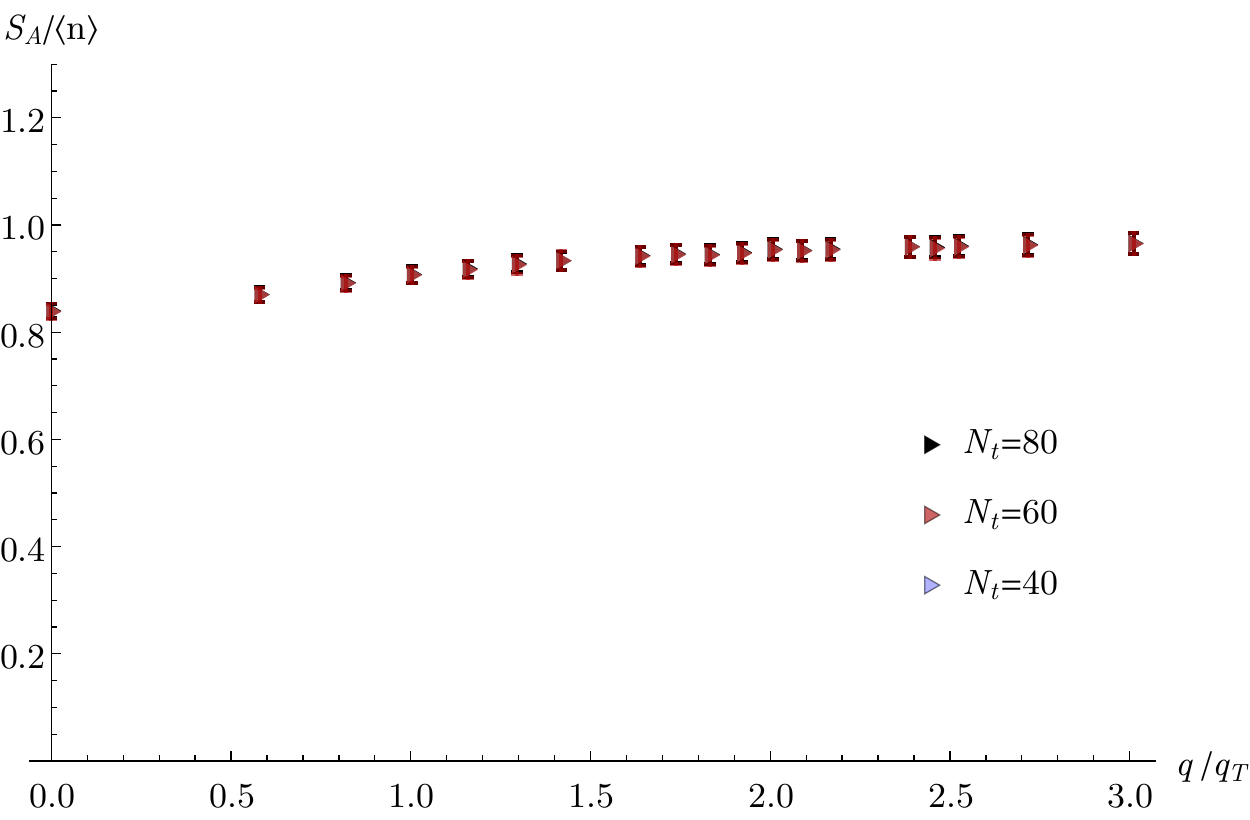}
\caption{Extrapolations of the vector (top row) and axial (bottom row) structure factors at $z=0.1$. The first column are infinte volume limits, the second column is the $\Delta x \rightarrow 0$ limit and the last column is the $\Delta t \rightarrow 0$ limit. There is little variation in the structure factors for all three limits.}
\label{fig:z0p1extrap}
\end{figure}

In order to explore the uncertainties arising from the finite volume and lattice spacing employed in our calculations we first compute the structure factors for two values of the fugacity, $z=0.1$ and $z=1.0$ using three sets of parameters (and $M=1$, $T=4.13$) shown in \tab{tab:extrapol-param}. 
\begin{table}[b]
\begin{center}
\begin{tabular}{|c| c| c|}
\hline
Volume extrapolation &  Space continuum extrapolation & Time continuum extrapolation \\
\hline
&&\\
$\Delta x  =0.4, \Delta t = 0.01, N_t=40,$  
&
$\Delta x = 0.4 \frac{5}{N_x},  \Delta t = 0.01, N_t=40, $  
& 
$\Delta x = 0.4 , N_x=7,  \Delta t = 0.01 \frac{40}{N_t}, N_t=40,$
 \\
 $N_x = 5,7,9$ &  $N_x = 5,7,9$ & $N_t = 40, 60, 80$\\
    \hline
\end{tabular}
\end{center}
\caption{Three sets of parameters used in finite volume, space and time continuum extrapolations.}
\label{tab:extrapol-param}
\end{table} We use lattices with odd number of points in the spatial direction because such lattices converge to the infinite volume limit faster than those with even number of sites. For nuclear astrophysics applications it is useful to think of the systems of units we use as corresponding to $\Delta x = 1 = 5 \fm, M=1=938 \MeV, \Delta t = 0.01 = 1.19 \fm$.
The results we find are shown on \fig{fig:z1p0extrap} and \fig{fig:z0p1extrap}. They show that the calculations performed on the smallest volume and coarsest lattices are already sufficient to determine the infinite volume and continuum results, within the statistical errors. The only exception are the results for the densest systems  in our range ($z=1.0$).
Only the two finest lattice results for $S_A$ ($\Delta x = 0.4 \times 5/7,  0.4 \times 5/9 $),  agree well within the statistical uncertainties while the results $S_V$ are different for all three values of $\Delta x$. Even in this case, however, signs of convergence are clear, with the difference between the $\Delta x = 0.4 $ and $\Delta x = 0.4 \times 5/7$ results being much smaller than between $\Delta x = 0.4 \times 5/7$ and $\Delta x = 0.4 \times 5/9$. Due to this dependence on $\Delta x$, we assign a systematic uncertainty of
$5\%$ on the results for the vector structure factors at the largest values of $z$.

\begin{figure}[t]
\includegraphics[width=8cm]{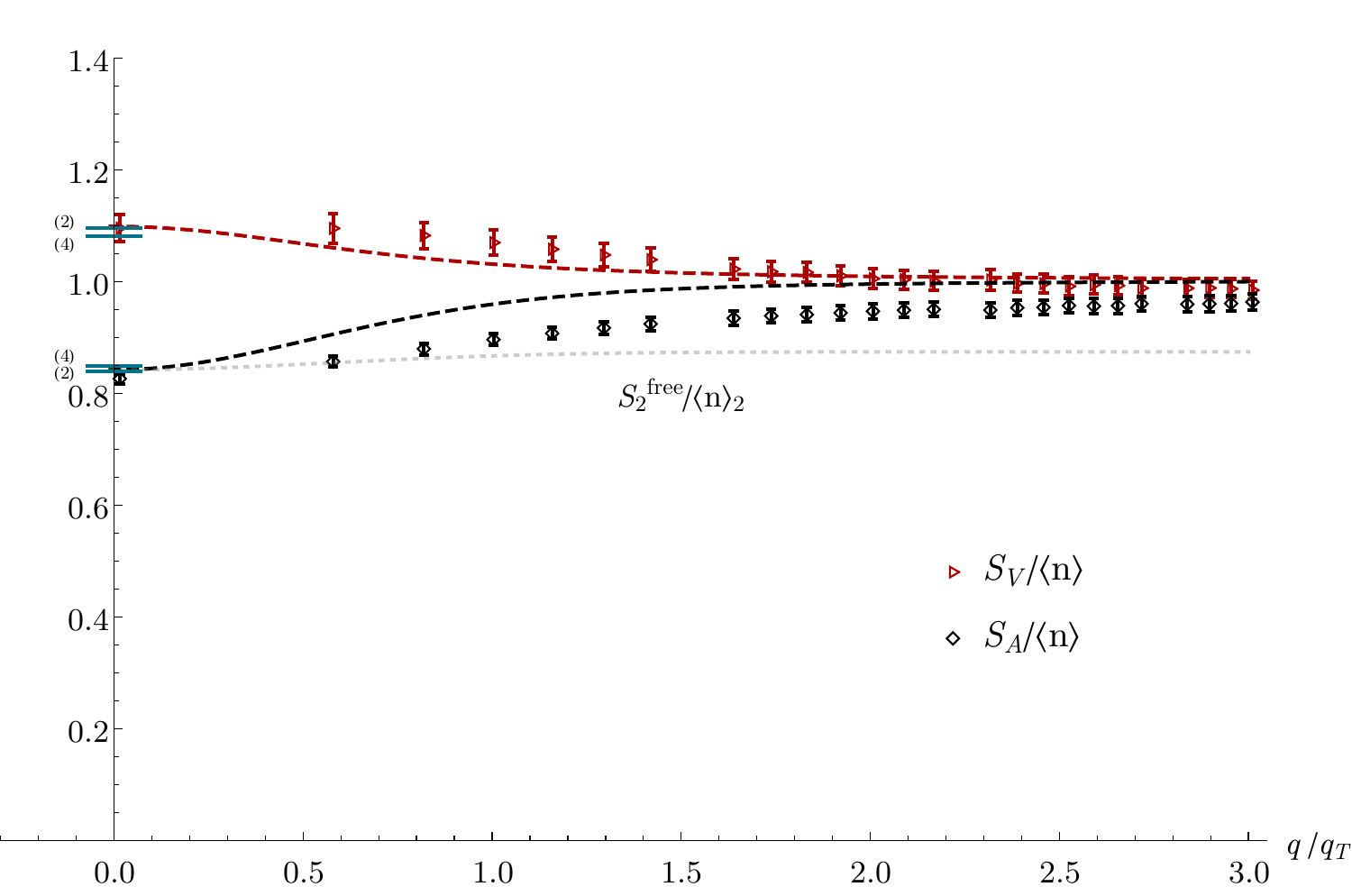}
\includegraphics[width=8cm]{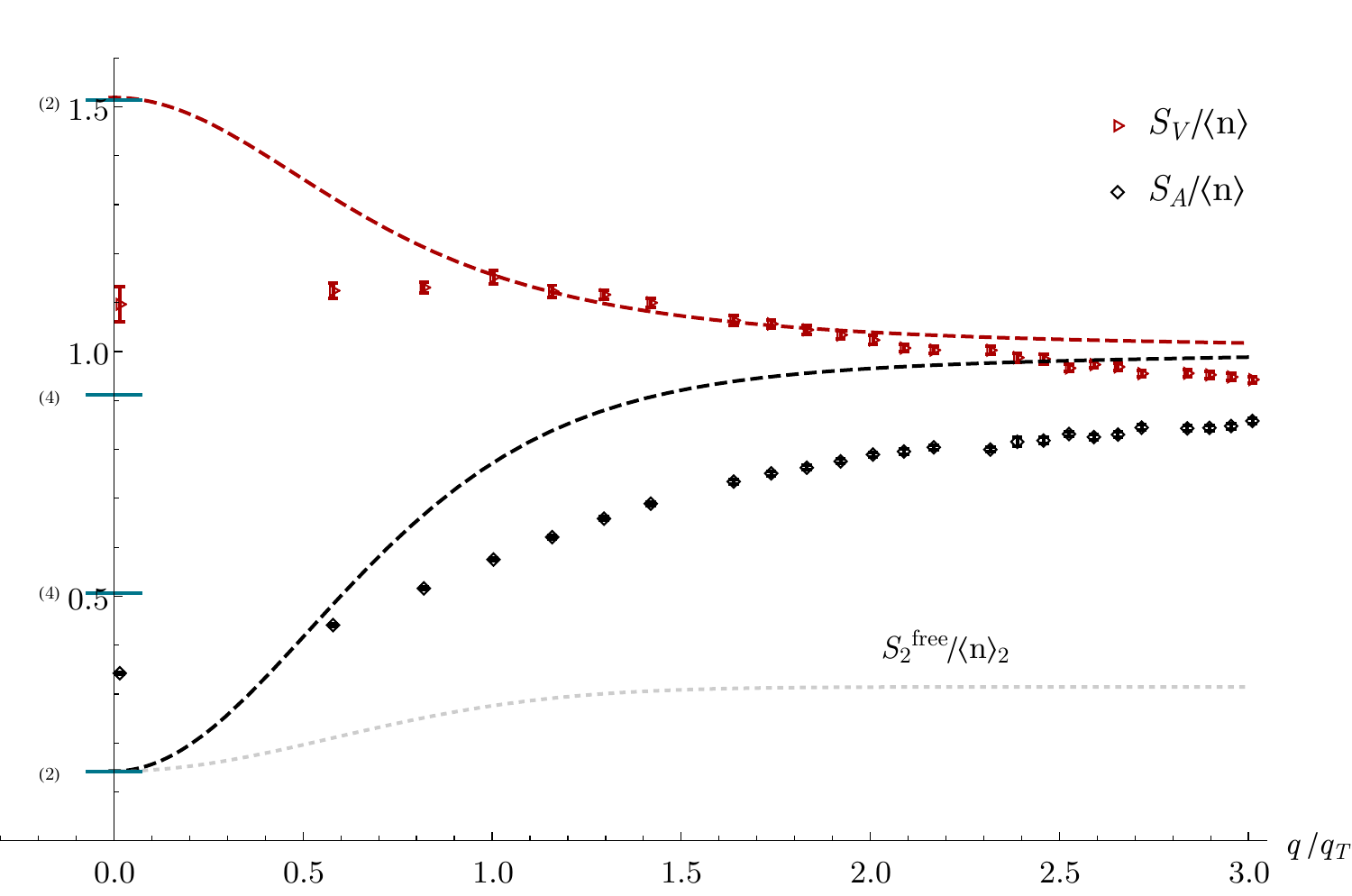}
\caption{Lattice results vs virial expansion calculations. We plot the scaled structure factors as a function of $q/q_T$ for $z=0.1$ (left) and $z=1.0$ (right). The red data is the vector structure factor and the black is the axial structure factor. The dashed curves are calculations using a combination of the virial expansion to second order plus the pseudopotential method \cite{PhysRevC.98.015802}. The dashed gray lines are the static structure factors computed in the free theory to $\mathcal{O}(z^2)$, scaled by the density computed to $\mathcal{O}(z^2)$. It is clear that, even at small $z$, interactions produce a sizeable correction to the structure factors. The short horizontal bands in blue at small $q/q_{T}$ are the various sum rules, calculated to $\mathcal{O}(z^2)$ and $\mathcal{O}(z^4)$ as discussed in the text.}
\label{fig:virial-comparison}
\end{figure}

The asymptotic limit of the structure factors are known on general grounds:
\beq
S_V(q\rightarrow \infty) = S_A(q\rightarrow \infty) = n\,.
\eeq 
Since we are limited by the lattice cutoff to values of $q< \pi/\Delta x$ we can see that this limit is reached in less dense system ($z=0.1$) but larger values of the momentum would be needed to verify it at more dense systems ($z=1.0$), as can be seen on \fig{fig:virial-comparison}.
In the same figure we compare our  results to some known analytic results. The $S_{V,A}(q\rightarrow 0)$ limits are determined by thermodynamics 
\beqs
S_V(q\rightarrow 0) &=T\Big(\frac{\partial n}{\partial \mu}\Big)\,, \\
S_A(q\rightarrow 0) &= T\Big(\frac{\partial s}{\partial h}\Big)\,. 
\eeqs
These have been computed  up to fourth order in the virial expansion \cite{PhysRevC.96.055804} and are shown in blue.
The second and fourth order virial result is shown as small segments  on the vertical axis. The second order in the virial expansion of a {\it free} gas can also be easily computed and is shown in \fig{fig:virial-comparison} as a grey dotted line. Virial results for non-zero value of the momentum are not yet available. However, a calculation up to second order in $z$ using a pseudo-potential interaction (tuned to unitarity) and finite values of $q$ is available \cite{PhysRevC.98.015802} and shown as dashed lines in \fig{fig:virial-comparison}  \footnote{The main motivation of the calculation in \cite{PhysRevC.98.015802} was the computation of {\it dynamical} structure factors which cannot be computed with Monte Carlo methods of the kind discussed in this paper. }.  As expected, all virial calculations agree well with our data at small 
($z\alt 0.5$) but differ significantly at larger values of $z$.

Finally, we have compared with a recent calculation by Jensen et al~\cite{Jensen:2018opr} of the thermodynamics of the unitary gas, including the $S_{A}(q\rightarrow 0) $ limit. The authors explore the parameter space $0.1 < T/T_F < 0.4$, where $T_F = (3 \pi^2 n)^{2/3}/2M$. 
We extended our calculations to lower temperatures to be able to compare with their values. 
Jensen et al. report 
$S_{A}(q\rightarrow 0)/S_0 =0.403(1) $ at $T/T_F=0.353$ (where $S_0 = \frac{3n}{2}\frac{T}{T_F}$) on $9^3$ lattices, while we  obtain $S_{A}(q\rightarrow 0)/S_0 =0.39(1) $  at $T/T_F=0.371(5) $. Our results therefore appear consistent with those of Jensen et al. Another calculation~\cite{PhysRevLett.110.090401} reports values about $20\%$ higher.

\begin{figure}
\includegraphics[width=0.49\textwidth]{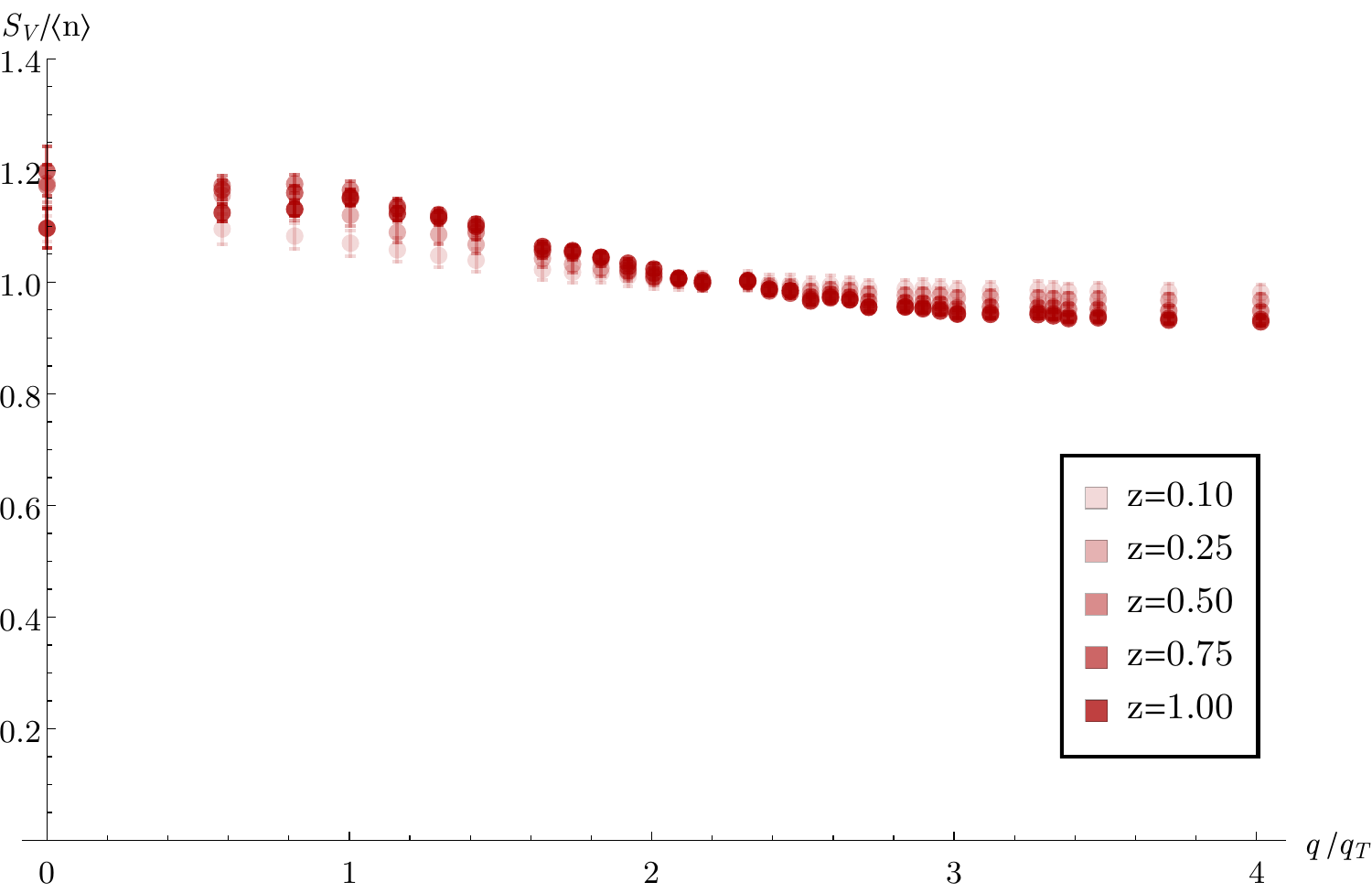}
\includegraphics[width=0.49\textwidth]{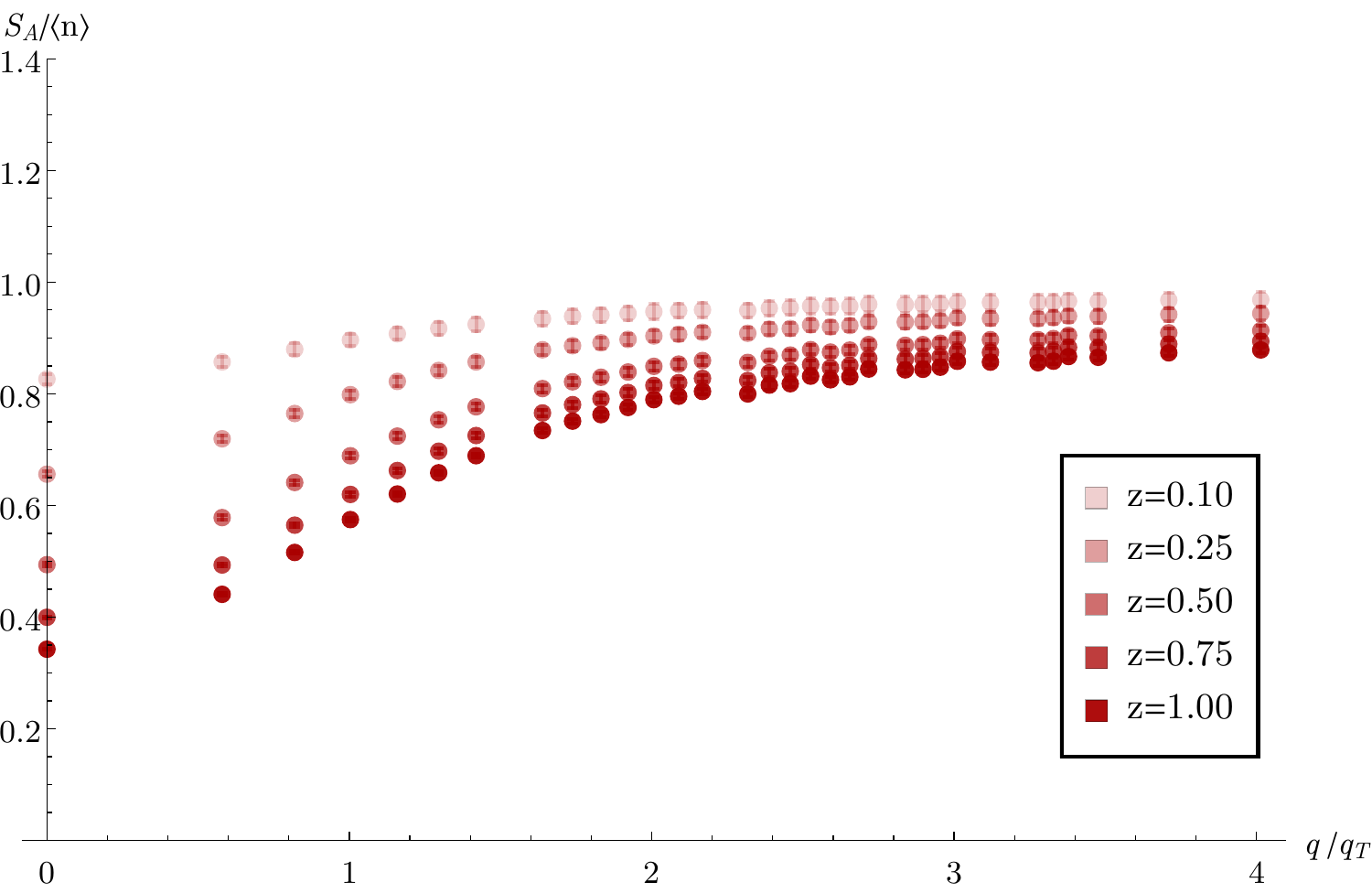}
\caption{Scaled structure factors for different fugacities as a function of the momentum in units of the thermal momentum $\sqrt{6MT}$.  }
\label{fig:z-sweep}
\end{figure}

Our main results are summarized in \fig{fig:z-sweep}. The error bar include the statistical errors but it should  be kept in mind that an additional systematic for the vector structure factor reaching up $5\%$ for the highest $z=1.0$ should be added to the uncertainties.

\section{conclusions} \label{conclusions}
We computed the static structure factors of the unitary gas in the $0.6 T_F \alt T \alt 6 T_F$ using Monte Carlo methods, corresponding to  fugacities in the  $0.1 \alt z \alt 1.0$ range. 
%The lattice formulation allows us to push beyond the virial regime, and we compute structure factors for fugacities $0.1 \leq z \leq 1.0$. Due to scaling laws arising from universality, though we compute at the single temperature of $T=4.13 \text{ MeV}$, our results can be used for arbitrary temperature above the superfluid phase transition by a simple, exact scaling law. 
The unitary gas at low densities is a reasonable model for  a dilute neutron gas and this range of fugacities is the one relevant to the  neutrinosphere (the region around a proto-neutron star where neutrinos decouple) in core collapse supernova. The structure factors are the relevant many-body information to describe the interactions between the neutrinos and the neuron matter that drives the explosion.
%Our primary purpose for calculating these structure factors is for use in supernova simulation codes, where in the neutrino decoupling region, in-medium neutral current scattering processes are determined to a very good approximation by the vector and axial static structure factors. The neutrinosphere is primarily composed of neutrons, which have large scattering ${}^1 S_0$ scattering lengths, and we expect the unitary gas to approximate the behavior of neutrons at low density.
%We find that our results satisfy various exact sum rules provided by thermodynamics and unitarity. 
We compared our lattice calculations of the static structure factors with an $\mathcal{O}(z^2)$ calculation using the pseudopotential approach. We find reasonable agreement with the second order virial expansion at $z=0.1$, while we find qualitative differences between the lattice and virial expansion at $z=1.0$. In particular, the dramatic bump in the vector structure factor at small $q/q_{T}$ predicted by the virial expansion is in fact not present.

We demonstrate that, for all densities explored, both the infinite volume and the $\Delta t\rightarrow 0$ limits are very well behaved, with negligible variation in both structure factors. At small density, the $\Delta x \rightarrow 0$ limit also yields negligible variation in the structure factors. At $z=1.0$, the densest system we consider, we see some variation in the structure factors at the coarsest lattice spacings. We expect, however, that our calculations are within $5\%$ of the exact result at our finest lattice spacing.

 We also leave for future work an extension of this calculation to hot neutron matter that will require a more realistic interaction between neutrons, at least one leading to finite scattering lengths and non-zero effective range. 
 We also leave for future work a  more careful $\Delta x\rightarrow 0$ extrapolations. Such extrapolations require calculations on larger lattices, which are more expensive, but there is no fundamental difficulty in doing so.
 %A calculation of the static structure factors for hot neutron matter will be useful in understanding the neutrinosphere, the region around a proto-neutron star where neutrinos decouple from the hot and dense nuclear matter created by a supernova explosion. This will require the tuning of the scattering length away from unitarity and the addition of an effective range. This would be the first exact calculation of the static structure factors of hot and dense neutron gases and will be useful for the astrophysics community by deepening our understanding of the supernova explosion mechanism.
 
\begin{acknowledgments}
A. A. is supported in part by the National Science Foundation CAREER grant PHY-1151648 and by U.S. Department of Energy under Contract No. DE-FG02-95ER-40907. A.A. acknowledges the hospitality of the University of Maryland where part of this work was performed.  P. B. and N.W. are supported by the U.S. Department of Energy under Contract No.~DE-FG02-93ER-40762. P.B. and N.W. acknowledge the hospitality of The George Washington University where part of this work was performed.
\end{acknowledgments}

\bibliographystyle{apsrev4-1}
\bibliography{unitary-supernova-bib}

\end{document}